\documentclass{raa}           
\usepackage{graphicx}
\usepackage{times}
\usepackage{natbib}
\usepackage{amssymb,amsmath}
\usepackage{multirow}
\bibpunct{(}{)}{;}{a}{}{,}
\usepackage{hyperref}
\hypersetup{pdftitle = Validation of LAMOST stellar parameters, pdfauthor = Hua Gao, pdfsubject= LAMOST stellar parameters, pdfkeywords = stellar
 parameters catalogs} 
\hypersetup{colorlinks = true, linkcolor = green, anchorcolor = red, citecolor = blue, filecolor = red, pagecolor = red, urlcolor = red}

\begin{document}
\title{Validation of LAMOST Stellar Parameters with the PASTEL Catalog
}

 \volnopage{ {\bf } Vol.\ {\bf X} No. {\bf XX}, 000--000}
   \setcounter{page}{1}

   \author{Hua~Gao\inst{1,2}, Hua-Wei~Zhang\inst{1,2}, Mao-Sheng~Xiang\inst{1,2}, Yang~Huang\inst{1,2}, Xiao-Wei~Liu\inst{1,2}, A-Li~Luo\inst{3},
     Hao-Tong~Zhang\inst{3}, Yue~Wu\inst{3}, Yong~Zhang\inst{4}, Guang-Wei~Li\inst{3}, Bing~Du\inst{3}}

\institute{Department of Astronomy, School of Physics, Peking University, Beijing 100871, China; {\it gaohua@pku.edu.cn, zhanghw@pku.edu.cn}\\
  \and Kavli Institute for Astronomy and Astrophysics, Peking University, Beijing 100871, China\\
  \and Key Laboratory of Optical Astronomy, National Astronomical Observatories, Chinese Academy of Sciences, Beijing 100012, China\\
  \and Nanjing Institute of Astronomical Optics \& Technology, National Astronomical Observatories, Chinese Academy of Sciences, Nanjing 210042, China\\
     \vs \no {\small Received ; accepted } }

   \abstract{Recently the Large Sky Area Multi-Object Fiber Spectroscopic Telescope (LAMOST) published its first data release (DR1), which is ranked
     as the largest stellar spectra dataset in the world so far. We combine the PASTEL catalog and SIMBAD radial velocities as a testing standard to
     validate the DR1 stellar parameters (effective temperature $T_{\mathrm{eff}}$, surface gravity $\log g$, metallicity $\mathrm{[Fe/H]}$ and radial
     velocity $V_{\mathrm{r}}$). Through cross-identification of the DR1 catalogs and the PASTEL catalog, we obtain a preliminary sample of 422
     stars. After removal of stellar parameter measurements from problematic spectra and applying effective temperature constraints to the sample, we
     compare the stellar parameters from DR1 with those from PASTEL and SIMBAD to prove that the DR1 results are reliable in restricted
     $T_{\mathrm{eff}}$ ranges. We derive standard deviations of 110 K, 0.19 dex, 0.11 dex and 4.91 $\mathrm{km\,s^{-1}}$ , for $T_{\mathrm{eff}}$,
     $\log g$, $[\mathrm{Fe/H}]$ when $T_{\mathrm{eff}}<8000\,\mathrm{K}$, and for $V_{\mathrm{r}}$ when $T_{\mathrm{eff}}<10000\,\mathrm{K}$,
     respectively. Systematic errors are negligible except for that of $V_{\mathrm{r}}$. Besides, metallicities in DR1 are systematically higher than
     those in PASTEL, in the range of PASTEL $[\mathrm{Fe/H}]<-1.5$.   } \authorrunning{H. Gao et.  al} 
   \titlerunning{Validation of LAMOST Stellar Parameters} 
   \maketitle

%
\section{Introduction}           
\label{sec:intro}
The formation and evolution of galaxies is one of the key astrophysical subjects at the moment. The Milky Way provides us a unique example to carry
out a detailed and comprehensive study. The Study of stars---the fundamental building blocks of galaxies, allows us to map the Galaxy and deepen our
understanding on galactic formation and evolution. Recent large stellar spectroscopic surveys such as the RAdial Velocity Experiment (RAVE;
\citealp{Steinmetz+RAVE+2003}), the Sloan Extension for Galactic Understanding and Exploration (SEGUE; \citealp{Yanny+SEGUE+2009}) and the APO
Galactic Evolution Experiment (APOGEE; \citealp{Allende_Prieto+APOGEE+2008}) have revolutionized our understanding of the Galaxy, though they all have
their emphases and limitations. Sizes of the stellar spectroscopic samples are limited due to galactic extinction and instrumental limitations. The
LAMOST, a Wang-Su Reflecting Schmidt Telescope located at the Xinglong Station of National Astronomical Observatories, Chinese Academy of Sciences
(NAOC), emerged in response to the pressing need for a more complete spectroscopic sample of the Galaxy \citep{Cui_the_LAMOST+2012}. Its large
aperture (effective aperture of $3.6$\textendash{}$4.9\,\mathrm{m}$) and wide field of view (20 square degrees) coexist, benefiting from the special
design that allows it to simultaneously obtain 4000 spectra from $3700$ to $9100\,\mathrm{\AA}$ in a single exposure at resolution $R\sim1800$.

The LAMOST Experiment for Galactic Understanding and Exploration (LEGUE) survey aims to provide a larger spectroscopic sample than ever before to
investigate kinematics and chemical abundances of the Galaxy \citep{Zhao+LM_survey_overview+2012,Deng+LEGUE+2012}. The survey consists of three
components: the spheroid survey, the Galactic anticenter survey and the disk survey. Each component has its own target selection strategy. The
spheroid survey will observe over 2.5 million stars in the North Galactic Cap and the South Galactic Cap selected from the Sloan Digital Sky Survey
(SDSS; \citealp{York_SDSS+2000}). The anticenter survey aims to sample the region $150\dg<l<210\dg$ and $-30\dg<b<30\dg$, with input targets selected
from the Xuyi photometric survey \citep{Zhang_Xuyi+2014}. The disk survey will cover the low latitude part $-20\dg<b<20\dg$ and will also use the Xuyi
photometric survey in conjunction with the third US Naval Observatory CCD Astrograph Catalog (UCAC3; \citealp{Zacharias_UCAC3+2010}) and the Two
Micron All Sky Survey (2MASS; \citealp{Skrutskie_2MASS+2006}) to prepare input targets. By the time our study is conducted, the LAMOST has produced
the largest stellar spectroscopic sample, which demonstrates the high efficiency of acquiring spectra endowed by the unique design.

After two years of test observation since the national acceptance in 2009, the LAMOST pilot survey began on 2011 October 24 and completed on 2012 June
24 \citep{Luo+pilot_survey+2012}. The subsequent regular survey started on 2012 September 28 and finished its first year mission on 2013 June 15. In
2013 August, DR1 of LAMOST became available to the Chinese astronomical community and international colleagues, including spectral products from the
pilot survey and the first-year regular survey \citep{Luo+DR1+2015}. The DR1 consists of 2,204,696 spectra of stars,
quasars, galaxies and some other objects of which the nature could not be established due to poor quality of their spectra.  To extract stellar
parameters from such large amount of spectra, the LAMOST stellar parameter pipeline (LASP) was developed \citep{Wu+ulyss+2011,Wu+LASP+2014}. It
employs the ULySS software \citep{Koleva+ulyss+2009} to analyze the LAMOST spectra and derives full set of stellar atmospheric parameters
($T_{\mathrm{eff}}, \log g, [\mathrm{Fe/H}]$) and radial velocities through minimizing the $\chi^{2}$ value between the observed spectrum and a model
spectrum generated by an interpolator built based on the ELODIE library \citep{Prugniel+ELODIE+2001,Prugniel+ELODIE1+2007}. Some of radial velocities
in DR1 are LASP measurements, and the rest are products of 1D pipeline \citep{Luo+pilot_survey+2012} when LASP measurements are not available. There
are 1,944,329 stellar spectra in DR1 catalogs, but only 1,061,918 of these spectra have yielded full set of stellar atmospheric parameters and radial
velocities due to quality control. Only late A or FGK type stars with $g$-band signal-to-noise ratio $\mathrm{S/N}\geq15$ or $\mathrm{S/N}\geq6$ for
bright and dark nights are allowed for LASP input, respectively \citep{Wu+LASP+2014}.

The reliability of these spectral products has to be investigated before any further applications. The PASTEL catalog provides a good testing
standard. It is a catalog of stellar atmospheric parameters of tens of thousands stars compiled by surveying bibliographies in the main astronomical
journals and the CDS database. Determination of most stellar parameters in the catalog is based on analysis of high-resolution and high-$\mathrm{S/N}$
spectra, though some recent precise $T_{\mathrm{eff}}$ measurements that are not based on high-resolution spectra are also included
\citep{Soubiran+the+PASTEL+2010}. The majority of the stars in the catalog are FGK stars. $90\%$ of the stars in the catalog have $V$ magnitude
brighter than $9.75\,\mathrm{mag}$. Internal errors of their parameters are $1.1\%$, $0.10\,\mathrm{dex}$ and $0.06\,\mathrm{dex}$ for
$T_{\mathrm{eff}}$, $\log g$ and $[\mathrm{Fe/H}]$, respectively. In addition, the catalog provides information such as the equatorial coordinates and
$B$, $V$, $J$, $H$, $K$ magnitudes retrieved from the SIMBAD database \citep{Wenger+SIMBAD+2000}, on which our cross-identification of common targets
is based.

We firstly perform a coarse cross-identification of the DR1 catalogs and the PASTEL catalog and then gradually refine the sample.  Detailed
information of the cross-identification sample and the subsequent refinements will be discussed in Section~\ref{sec:cro_res}. Subsequently we compare
the stellar parameters from the DR1 catalogs with those from the PASTEL catalog and the SIMBAD database. Results of the comparison are shown and
discussed in Section~\ref{sec:com_res}. Finally, we summarize our work in Section~\ref{sec:sum}.

\section{Cross-identification and the Sample for Validation}
\label{sec:cro_res}

\subsection{Cross-identification}
\label{sec:cr}

In order to compare stellar parameters from DR1 and PASTEL, we first identify common objects in the DR1 catalogs and the PASTEL catalog. The PASTEL
catalog is regularly updated. The version that we use is 17-May-2013, which consists of 52,045 entries of 26,657 individual stars. Due to the position
offsets of the bright stars observed with LAMOST \citetext{H.-T. Zhang, private communication}, we expand search radius to 10$\,$arcsec.  We
carefully check the common entries given by the loose position criteria according to consistency of the photometry data in the DR1 catalogs and the
PASTEL catalog and images of targets from SIMBAD database. After removal of false positives, we obtain a raw sample of 422 stars.

\subsection{Stellar Parameters in the Sample}
\label{sec:ste_param}

Stellar atmospheric parameters and radial velocities of more than 300 stars in the sample are available in the A, F, G and K type stars catalogs of
DR1.  Only radial velocity measurements are available in DR1 catalogs for the remaining stars. A total of 420 stars in the sample have
$T_{\mathrm{eff}}$ measurements in the PASTEL catalog, while $\log g$ and $\mathrm{[Fe/H]}$ are available for about only 150 stars in the sample.
That makes us short of testing standards for DR1 $\log g$ and $\mathrm{[Fe/H]}$ measurements. We retrieve radial velocities of 323 stars in the sample
from SIMBAD database for further analysis.

For some PASTEL stars that have multiple measurements, we remove the very old (published before 1990) measurements that deviate from the more recent
ones, and we average the remaining measurements as adopted values. Thus, each star in the comparison sample has a unique testing standard. For DR1
stars we retain every entry when multiple measurements are available in the catalogs.  We notice that some of DR1 stellar parameters show great
deviation from their testing standards through tentative comparison. We select these potential outliers with the criteria
$|T_{\mathrm{eff}}(\mathrm{DR1})-T_{\mathrm{eff}}(\mathrm{PASTEL})|\geq 1000\,\mathrm{K}$,
$|\log g(\mathrm{DR1})-\log g(\mathrm{PASTEL})|\geq 0.5\,\mathrm{dex}$,
$|[\mathrm{Fe/H}](\mathrm{DR1})-[\mathrm{Fe/H}](\mathrm{PASTEL})|\geq 0.5\,\mathrm{dex}$ or
$|V_{\mathrm{r}}(\mathrm{DR1})-V_{\mathrm{r}}(\mathrm{SIMBAD})|\geq 20\,\mathrm{km\,s^{-1}}$ for further inspection. We retrieve DR1 spectra of
these potential outliers for careful examination. We find some of these spectra were obtained under poor observation conditions, which might have
caused problems in determination of the stellar parameters. Spectra with low $\mathrm{S/N}$ (i.e. $\mathrm{S/N}<7$) yield poor estimates of stellar
parameters. Some spectra with bad pixel masks leave little information for stellar parameter measurements, therefore results from these broken spectra
are not reliable. In addition to the quality of spectra, we examine the validity of cross-identification of these potential outliers. We find that
some stars residing in binaries or star clusters are prone to causing the problem of fiber mis-pointing.  In consideration of position offset
($\sim4$-$9''$) during LAMOST observation, these stellar parameters may belong to another star in the crowded field, which is responsible for dramatic
deviation of the stellar parameters. We also discover a few mistakes in the PASTEL catalog by checking the original bibliographies.  The catalog has
mistaken stellar parameters of KIC 5524720 for those of TYC 3125-2594-1, stellar parameters of TYC 2667-624-1 for those of TYC 2267-624-1 and stellar
parameters of SAO 201781 for those of HD 201781.

\begin{figure}
  \centering
  \includegraphics[width=0.7\textwidth]{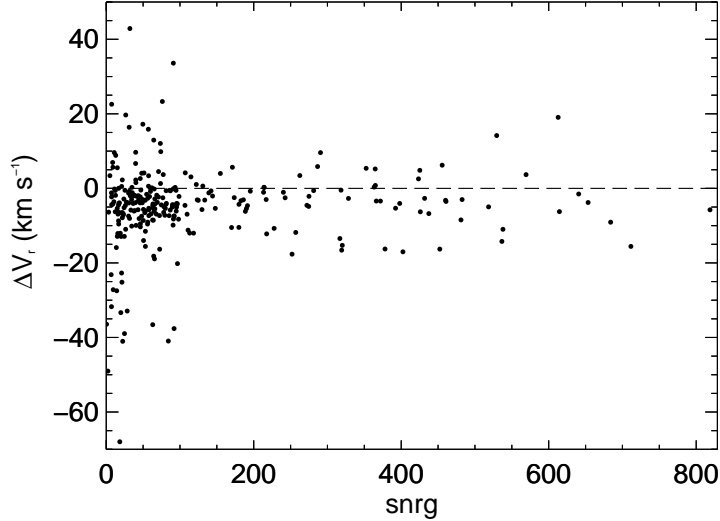}
  \caption{Precisions of DR1 $V_{\mathrm{r}}$ as compared with those from the SIMBAD database depend on the $g$-band $\mathrm{S/N}$, where
      $\triangle V_{\mathrm{r}}$ is defined to be $V_{\mathrm{r}}(\mathrm{DR1})-V_{\mathrm{r}}(\mathrm{SIMBAD})$, and $snrg$ is $g$-band
      $\mathrm{S/N}$ extracted from the DR1 catalogs.}
  \label{fig:1}
\end{figure}

Before we remove all these outliers, it is necessary to investigate how the $\mathrm{S/N}$ affects precisions of DR1 stellar parameters. We remove all
outliers mentioned above except for the low-$\mathrm{S/N}$ ones. The fact that majority of the sample are bright stars results in a lack of
low-$\mathrm{S/N}$ statistics that are required to investigate dependence of parameter precisions on the $\mathrm{S/N}$, which is obvious in
Figure~\ref{fig:1}. $85\%$ of the DR1 measurements have $g$-band $\mathrm{S/N}\geq20$, for those $V_{\mathrm{r}}$ testing standards are available in
the SIMBAD database. It is unrealistic to draw conclusions about which $\mathrm{S/N}$ range our following comparison results will hold. However, we
expect that for a sample with lower $\mathrm{S/N}$, derived precisions of DR1 measurements will be poorer. Finally, we remove the remaining
low-$\mathrm{S/N}$ ($\mathrm{S/N}<7$) outliers as well.

\begin{figure}
  \centering
  \includegraphics[width=1\textwidth]{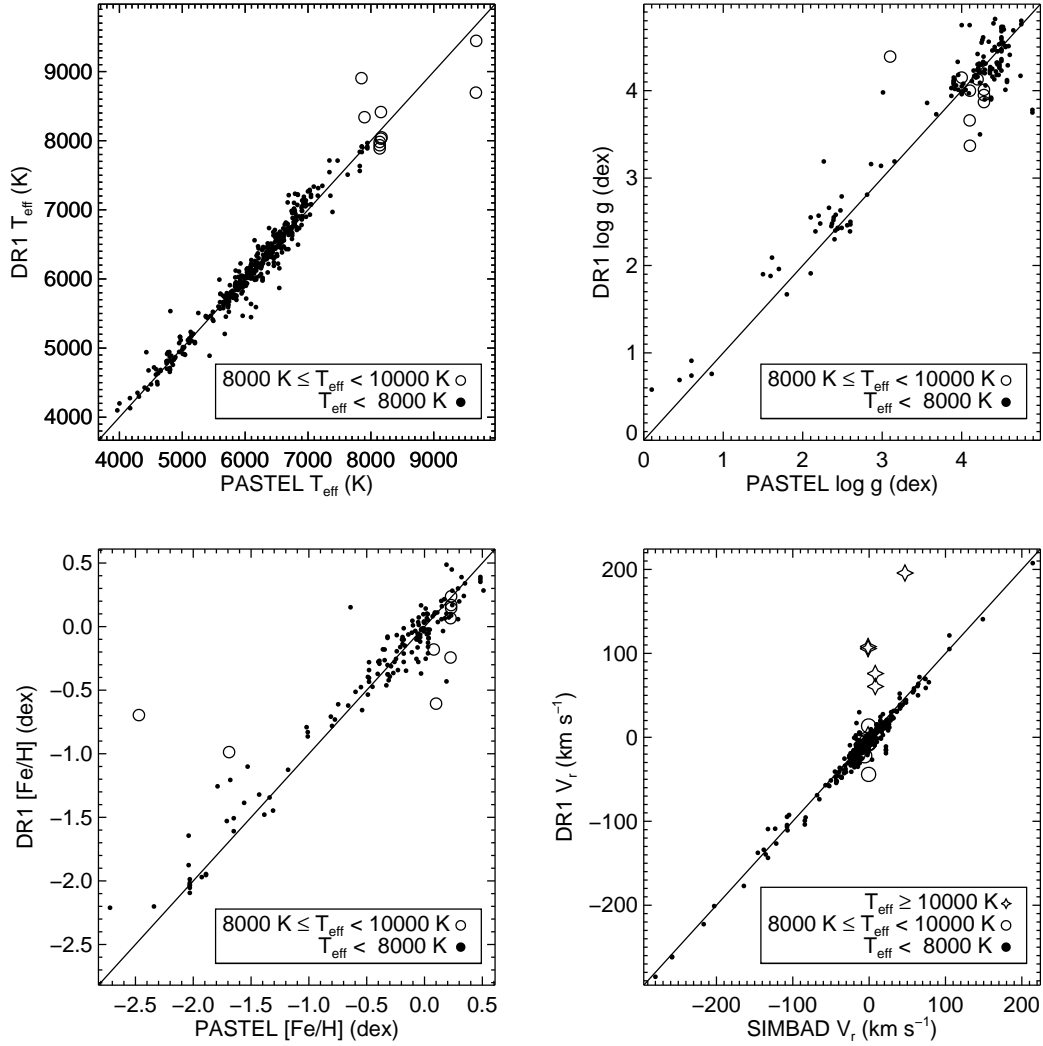}
  \caption{Tentative comparison of stellar parameters from DR1 with those from PASTEL and SIMBAD after discarding problematic measurements. Filled
    circles indicate measurements in the $T_{\mathrm{eff}}<8000\,\mathrm{K}$ bin, open circles represent measurements in the
    $8000\,\mathrm{K}\leq T_{\mathrm{eff}}<10000\,\mathrm{K}$ bin and open stars for measurements in the $T_{\mathrm{eff}}\geq10000\,\mathrm{K}$ bin.}
  \label{fig:2}
\end{figure}

Discarding problematic spectral products excludes impacts from other factors, and highlights how precisions vary for different spectral types of
stars. We group the DR1 measurements into three effective temperature bins: $T_{\mathrm{eff}}<8000\,\mathrm{K}$,
$8000\,\mathrm{K}\leq T_{\mathrm{eff}}<10000\,\mathrm{K}$ and $T_{\mathrm{eff}}\geq10000\,\mathrm{K}$. In order to avoid ambiguity, we use different
grouping strategies for different bins; we group the DR1 measurements into the $T_{\mathrm{eff}}<8000\,\mathrm{K}$ bin only when both their
corresponding PASTEL effective temperatures and the DR1 effective temperatures satisfy $T_{\mathrm{eff}}<8000\,\mathrm{K}$; for the other two bins, we
accept measurements when their corresponding PASTEL effective temperatures or the DR1 effective temperatures fall within the temperature ranges. Since
there are only a handful of measurements in the two high temperature bins, we manually examine each of them and find that there is no duplicate
occurrence in these two bins. We find that most DR1 stellar atmospheric parameters of stars hotter than $8000\,\mathrm{K}$ show great deviation from
the testing standards (see Fig.~\ref{fig:2}).  The $V_{\mathrm{r}}$ measurements in the $8000\,\mathrm{K}\leq T_{\mathrm{eff}}<10000\,\mathrm{K}$ bin
still show moderate accuracy as compared with the SIMBAD radial velocities. Finally, we confine our sample to an effective temperature range
$T_{\mathrm{eff}}<8000\,\mathrm{K}$ for validation of $T_{\mathrm{eff}}$, $\log g$, $[\mathrm{Fe/H}]$, and another effective temperature range
$T_{\mathrm{eff}}<10000\,\mathrm{K}$ for validation of $V_{\mathrm{r}}$, where $T_{\mathrm{eff}}$ here stands for effective temperature of both DR1
and PASTEL. 
\subsection{Internal Scatter of the DR1 Stellar Parameters}
\label{sec:internal_scatter}

\begin{figure}
  \centering
  \includegraphics[width=0.85\textwidth]{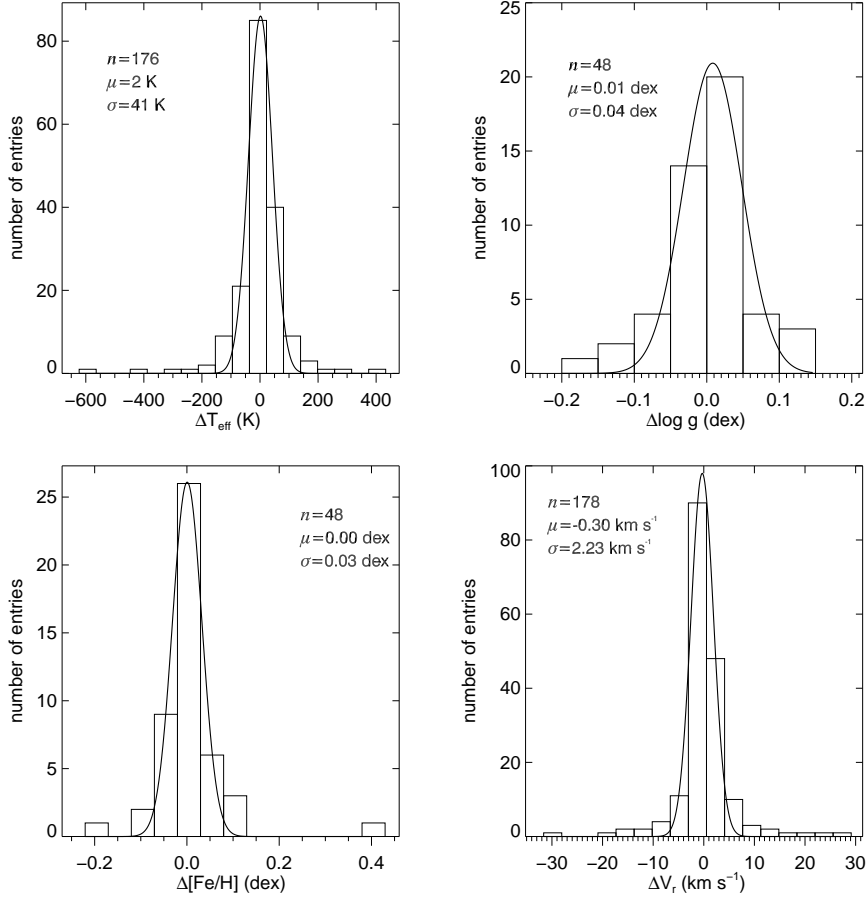}
  \caption{Internal scatter of the DR1 multiple measurements of stellar parameters. Internal scatter  is defined by
    $\triangle P=P_{\mathrm{mul}}-P_{\mathrm{adop}}$, where $P$ is one of the four stellar parameters, ``mul'' stands for multiple measurements in DR1
    and ``adop'' represents adopted ``mean'' value of the multiple measurements. In all panels, $n$ is number of multiple measurements,
      $\mathrm{\mu}$ is mean of the Gaussian fit and $\mathrm{\sigma}$ is the standard deviation of the Gaussian fit.}
  \label{fig:3}
\end{figure}

Now that we have obtained a clean sample, the next step is to combine the multiple measurements in the DR1 catalogs for individual stars which have
corresponding testing standards available in the PASTEL catalog and in the SIMBAD database. When the star was observed more than once, we remove
relatively low-$\mathrm{S/N}$ measurements that show significant offsets from the others, and then adopt the mean value of the remaining measurements.
Note that there were only a few low-$\mathrm{S/N}$ measurements left after we removed those outliers in Section~\ref{sec:ste_param}. We retrieve 176
multiple $T_{\mathrm{eff}}$ measurements for 73 stars, 48 multiple $\log g$ measurements for 19 stars, 48 multiple $\mathrm{[Fe/H]}$ measurements for
19 stars and 178 multiple $V_{\mathrm{r}}$ measurements for 73 stars to validate our choice of their ``mean'' values.  The DR1 internal scatters of
multiple measurements are shown in Figure~\ref{fig:3} by taking the adopted stellar parameters as fiducials.  The internal errors as measured using
standard deviations of Gaussian fits are $41\,\mathrm{K}$, $0.04\,\mathrm{dex}$, $0.03\,\mathrm{dex}$ and $2.23\,\mathrm{km\,s^{-1}}$, for
$T_{\mathrm{eff}}$, $\log g$, $[\mathrm{Fe/H}]$ and $V_{\mathrm{r}}$, respectively. Our adopted stellar parameters appear to be reasonable ``mean''
values of multiple measurements since there are only a few measurements that show great deviation from the adopted value.
\section{Comparison of stellar parameters}
\label{sec:com_res}

\begin{figure}
  \centering
  \includegraphics[width=13cm,height=22cm]{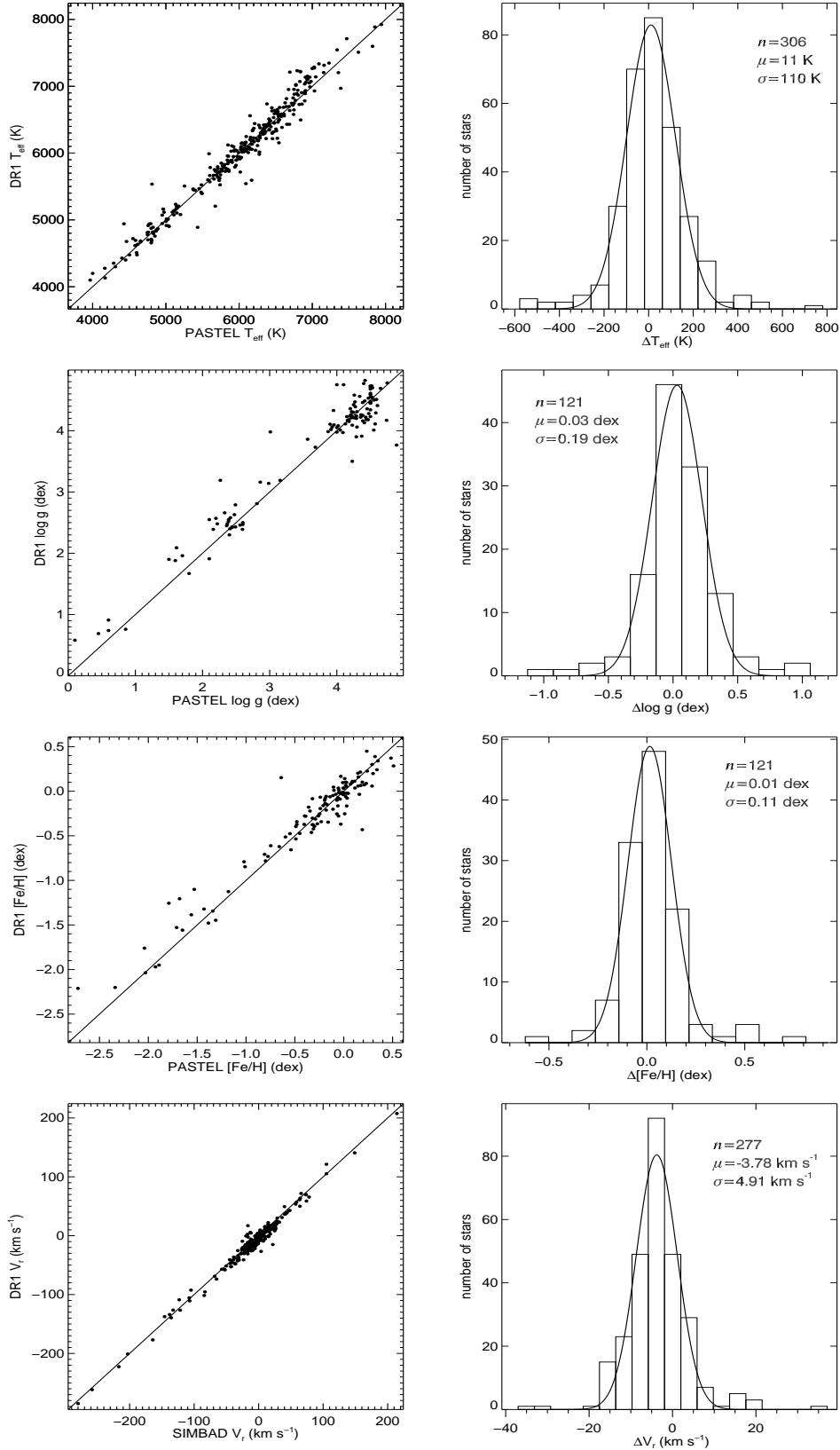}
  \caption{Comparison of stellar parameters from DR1 with those from the PASTEL catalog and the SIMBAD database for individual stars in the clean
    sample. The offsets of the stellar parameters are defined as $\triangle P=P(\mathrm{DR1})-P(\mathrm{PASTEL/SIMBAD})$, where $P$ is one of the four
    stellar parameters. In the four panels of the right column, $n$ is number of stars, $\mathrm{\mu}$ is mean of the Gaussian fit and
    $\mathrm{\sigma}$ is the standard deviation of the Gaussian fit.}
  \label{fig:4}
\end{figure}

After removal of spectral products from poor quality spectra and misidentified entries, we applied different temperature constraints to the sample,
because the DR1 $V_{\mathrm{r}}$ measurements have a good enough accuracy over the high temperature range. Eventually, after combining multiple DR1
measurements, we derive a clean and well-established sample of 306 stars for $T_{\mathrm{eff}}$ comparison, 121 stars for $\log g$ comparison, 121
stars for $\mathrm{[Fe/H]}$ comparison and 277 stars for $V_{\mathrm{r}}$ comparison. Figure~\ref{fig:4} displays, from top to bottom, the comparison
results of $T_{\mathrm{eff}}$, $\log g$, $[\mathrm{Fe/H}]$ and $V_{\mathrm{r}}$, respectively. Systematic errors and standard deviations listed in
Table~\ref{tab:1} are calculated through Gaussian fits.

Systematic errors of stellar parameters are negligible except for that of $V_{\mathrm{r}}$. Since radial velocities in the SIMBAD database are
collections from different literature, systematic biases of these measurements are expected to have canceled out. Thus, a systematic error of
$-3.78\,\mathrm{km\,s^{-1}}$ should be taken into account when $V_{\mathrm{r}}$ being used. We also find that DR1 metallicities of metal-poor stars
(i.e. PASTEL $\mathrm{[Fe/H]}<-1.5\, \mathrm{dex}$) are systematically higher than metallicities obtained from high-resolution spectra analysis;
statistic of 12 stars in this region gives a overestimation of $0.23\,\mathrm{dex}$. Stellar parameters from DR7 of SDSS also show similar behavior
\citep{Xu+EMP+2013}. One should use caution when using these derived metallicities. Metallic lines for these metal-poor stars are so weak that
spectral noise could dominate over these spectral regions. We speculate that such overestimation might me caused by the significant spectral noise
somehow.

\citet{Lee+SSPP1+2008} showed that in an effective temperature range $4500\,\mathrm{K}\leq T_{\mathrm{eff}}\leq7500\,\mathrm{K}$, precisions of
stellar atmospheric parameters derived by the SEGUE Stellar Parameter Pipeline (SSPP; \citealp{Beers+SSPP0+2012}) are $141\,\mathrm{K}$,
$0.23\,\mathrm{dex}$, $0.23\,\mathrm{dex}$ for $T_{\mathrm{eff}}$, $\log g$ and $[\mathrm{Fe/H}]$, respectively, based on a comparison with analysis
of high-resolution spectra.  These statistics are comparable to the standard deviations listed in Table~\ref{tab:1}. These similarities are not
surprising because both SEGUE and LEGUE are medium-resolution spectroscopy surveys, and in addition their techniques used to derive stellar parameters
share some similarities such as template matching methods. DR1 has achieved similar precisions but provided a much larger dataset compared with SEGUE.

\begin{table}
\bc
\begin{minipage}[]{96mm}
  \caption{Statistics of Comparison Results for Individual Stars\label{tab:1}}
\end{minipage}
\begin{tabular}{c|cccc}
  \hline
  &$T_{\mathrm{eff}}$&$\log g$&$[\mathrm{Fe/H}]$&$V_{\mathrm{r}}$\\
  \hline
  Number of stars&306&121&121&277\\
  Systematic error&$11\,\mathrm{K}$&$0.03\,\mathrm{dex}$&$0.01\,\mathrm{dex}$&$-3.78\,\mathrm{km\,s^{-1}}$\\
  Standard deviation&$110\,\mathrm{K}$&$0.19\,\mathrm{dex}$&$0.11\,\mathrm{dex}$&$4.91\,\mathrm{km\,s^{-1}}$\\
  \hline
\end{tabular}
\ec
\end{table}
\section{Summary}
\label{sec:sum}

We perform cross-identification of the DR1 catalogs and the PASTEL catalog.  We set a search radius of 10$\,$arcsec to avoid missing the bright stars
with position offset.  We then remove false positives and obtain a preliminary sample of 422 stars. With this sample we make a tentative comparison
and select some potential outliers for further inspection. Finally, we discard results from problematic spectra, risky cross-identifications and a few
misidentifications in the PASTEL catalog itself. We obtain an effective temperature range $T_{\mathrm{eff}}<8000\,\mathrm{K}$ for validation of
$T_{\mathrm{eff}}$, $\log g$, $[\mathrm{Fe/H}]$, and another effective temperature range $T_{\mathrm{eff}}<10000\,\mathrm{K}$ for validation of
$V_{\mathrm{r}}$, because we do not expect that the DR1 measurements have equally good precisions in all effective temperature ranges. We derive
standard deviations of 110 K, 0.19 dex, 0.11 dex and 4.91 $\mathrm{km\,s^{-1}}$ for $T_{\mathrm{eff}}$, $\log g$, $[\mathrm{Fe/H}]$ when
$T_{\mathrm{eff}}<8000\,\mathrm{K}$, and for $V_{\mathrm{r}}$ when $T_{\mathrm{eff}}<10000\,\mathrm{K}$, respectively. The DR1 stellar parameters show
no systematic offsets except for $V_{\mathrm{r}}$ and overestimates of $\mathrm{[Fe/H]}$ at the low metallicity tail. We are looking forward to more
observations of PASTEL stars by LAMOST in the future, which provides us a larger sample to test the accuracy of LAMOST stellar parameters.
\begin{acknowledgements}
  We thank the anonymous referee for an expeditious review of this manuscript. H.G. and H.-W.Z. are greatful to Thijs Kouwenhoven, who helped to
  greatly improve the manuscript. This work was supported by the National Key Basic Research Program of China (NKBRP) 2014CB845700. This work was also
  supported by \nsfc{} (Grants No. 11473001 and 11233004). The Guoshoujing Telescope (the Large Sky Area Multi-Object Fiber Spectroscopic Telescope
  LAMOST) is a National Major Scientific Project built by the Chinese Academy of Sciences. Funding for the project has been provided by the National
  Development and Reform Commission. LAMOST is operated and managed by the National Astronomical Observatories, Chinese Academy of Sciences. This
  research has made use of the SIMBAD database, operated at CDS, Strasbourg, France.
\end{acknowledgements}
  
\bibliographystyle{raa}
\bibliography{ms2099}

\begin{thebibliography}{22}
\providecommand{\natexlab}[1]{#1}
\providecommand{\selectlanguage}[1]{\relax}

\bibitem[{{Allende Prieto} et~al.(2008){Allende Prieto}, {Majewski}, {Schiavon}
  et~al.}]{Allende_Prieto+APOGEE+2008}
{Allende Prieto}, C., {Majewski}, S.~R., {Schiavon}, R., et~al. 2008,
  Astronomische Nachrichten, 329, 1018

\bibitem[{{Beers} \& {Lee}(2012)}]{Beers+SSPP0+2012}
{Beers}, T., \& {Lee}, Y.~S. 2012, in Nuclei in the Cosmos (NIC XII), 94

\bibitem[{{Cui} et~al.(2012){Cui}, {Zhao}, {Chu} et~al.}]{Cui_the_LAMOST+2012}
{Cui}, X.-Q., {Zhao}, Y.-H., {Chu}, Y.-Q., et~al. 2012, Research in Astronomy
  and Astrophysics, 12, 1197

\bibitem[{{Deng} et~al.(2012){Deng}, {Newberg}, {Liu} et~al.}]{Deng+LEGUE+2012}
{Deng}, L.-C., {Newberg}, H.~J., {Liu}, C., et~al. 2012, Research in Astronomy
  and Astrophysics, 12, 735

\bibitem[{{Koleva} et~al.(2009){Koleva}, {Prugniel}, {Bouchard}, \&
  {Wu}}]{Koleva+ulyss+2009}
{Koleva}, M., {Prugniel}, P., {Bouchard}, A., \& {Wu}, Y. 2009, \aap, 501, 1269

\bibitem[{{Lee} et~al.(2008){Lee}, {Beers}, {Sivarani} et~al.}]{Lee+SSPP1+2008}
{Lee}, Y.~S., {Beers}, T.~C., {Sivarani}, T., et~al. 2008, \aj, 136, 2022

\bibitem[{{Luo} et~al.(2012){Luo}, {Zhang}, {Zhao}
  et~al.}]{Luo+pilot_survey+2012}
{Luo}, A.-L., {Zhang}, H.-T., {Zhao}, Y.-H., et~al. 2012, Research in Astronomy
  and Astrophysics, 12, 1243

\bibitem[{{Luo} et~al.(2015){Luo}, {Zhao}, {Zhao} et~al.}]{Luo+DR1+2015}
{Luo}, A.-L., {Zhao}, Y.-H., {Zhao}, G., et~al. 2015, ArXiv e-prints

\bibitem[{{Prugniel} \& {Soubiran}(2001)}]{Prugniel+ELODIE+2001}
{Prugniel}, P., \& {Soubiran}, C. 2001, \aap, 369, 1048

\bibitem[{{Prugniel} et~al.(2007){Prugniel}, {Soubiran}, {Koleva}, \& {Le
  Borgne}}]{Prugniel+ELODIE1+2007}
{Prugniel}, P., {Soubiran}, C., {Koleva}, M., \& {Le Borgne}, D. 2007, ArXiv
  Astrophysics e-prints

\bibitem[{{Skrutskie} et~al.(2006){Skrutskie}, {Cutri}, {Stiening}
  et~al.}]{Skrutskie_2MASS+2006}
{Skrutskie}, M.~F., {Cutri}, R.~M., {Stiening}, R., et~al. 2006, \aj, 131, 1163

\bibitem[{{Soubiran} et~al.(2010){Soubiran}, {Le Campion}, {Cayrel de Strobel},
  \& {Caillo}}]{Soubiran+the+PASTEL+2010}
{Soubiran}, C., {Le Campion}, J.-F., {Cayrel de Strobel}, G., \& {Caillo}, A.
  2010, \aap, 515, A111

\bibitem[{{Steinmetz}(2003)}]{Steinmetz+RAVE+2003}
{Steinmetz}, M. 2003, in GAIA Spectroscopy: Science and Technology,
  \emph{Astronomical Society of the Pacific Conference Series}, vol. 298,
  edited by U.~{Munari}, 381

\bibitem[{{Wenger} et~al.(2000){Wenger}, {Ochsenbein}, {Egret}
  et~al.}]{Wenger+SIMBAD+2000}
{Wenger}, M., {Ochsenbein}, F., {Egret}, D., et~al. 2000, \aaps, 143, 9

\bibitem[{{Wu} et~al.(2014){Wu}, {Luo}, {Du}, {Zhao}, \& {Yuan}}]{Wu+LASP+2014}
{Wu}, Y., {Luo}, A., {Du}, B., {Zhao}, Y., \& {Yuan}, H. 2014, ArXiv e-prints

\bibitem[{{Wu} et~al.(2011){Wu}, {Luo}, {Li} et~al.}]{Wu+ulyss+2011}
{Wu}, Y., {Luo}, A.-L., {Li}, H.-N., et~al. 2011, Research in Astronomy and
  Astrophysics, 11, 924

\bibitem[{{Xu} et~al.(2013){Xu}, {Zhang}, \& {Liu}}]{Xu+EMP+2013}
{Xu}, S.-Y., {Zhang}, H.-W., \& {Liu}, X.-W. 2013, Research in Astronomy and
  Astrophysics, 13, 313

\bibitem[{{Yanny} et~al.(2009){Yanny}, {Newberg}, {Johnson}
  et~al.}]{Yanny+SEGUE+2009}
{Yanny}, B., {Newberg}, H.~J., {Johnson}, J.~A., et~al. 2009, \apj, 700, 1282

\bibitem[{{York} et~al.(2000){York}, {Adelman}, {Anderson}
  et~al.}]{York_SDSS+2000}
{York}, D.~G., {Adelman}, J., {Anderson}, J.~E., Jr., et~al. 2000, \aj, 120,
  1579

\bibitem[{{Zacharias} et~al.(2010){Zacharias}, {Finch}, {Girard}
  et~al.}]{Zacharias_UCAC3+2010}
{Zacharias}, N., {Finch}, C., {Girard}, T., et~al. 2010, \aj, 139, 2184

\bibitem[{{Zhang} et~al.(2014){Zhang}, {Liu}, {Yuan} et~al.}]{Zhang_Xuyi+2014}
{Zhang}, H.-H., {Liu}, X.-W., {Yuan}, H.-B., et~al. 2014, Research in Astronomy
  and Astrophysics, 14, 456

\bibitem[{{Zhao} et~al.(2012){Zhao}, {Zhao}, {Chu}, {Jing}, \&
  {Deng}}]{Zhao+LM_survey_overview+2012}
{Zhao}, G., {Zhao}, Y.-H., {Chu}, Y.-Q., {Jing}, Y.-P., \& {Deng}, L.-C. 2012,
  Research in Astronomy and Astrophysics, 12, 723

\end{thebibliography}

\end{document}